\documentclass[vecphys]{svmult}

\usepackage{amsmath}
\usepackage{amssymb}

\usepackage{makeidx}         
\usepackage{graphicx}        
\usepackage{multicol}        
\usepackage[bottom]{footmisc}

\makeindex             

\begin{document}

\title*{Tessellations and Pattern Formation in Plant Growth and Development}
 \titlerunning{Patterns in Plant Growth}

\author{
Bruce E. Shapiro\inst{1}, 
Henrik J\"{o}nsson\inst{2},
Patrik Sahlin\inst{2},
Marcus Heisler\inst{3}, 
Adrienne Roeder\inst{3}, 
Michael Burl\inst{4}, 
Elliot M. Meyerowitz\inst{3} \and 
Eric D. Mjolsness\inst{5}\\
}
\authorrunning{Shapiro, B.E., et al} 

\institute{
Biological Network Modeling Center, Beckman Institute, California Institute of Technology, Pasadena CA, USA \texttt{bshapiro@caltech.edu}
\and 
Computational Biology and Biological Physics, Department of Theoretical Physics,  Lund University, Sweden
 \and
 Division of Biology, California Institute of Technology, Pasadena, CA, USA
 \and
 Jet Propulsion Lab, California Institute of Technology, Pasadena, CA, USA
 \and
 Department of Information and Computer Science and Institute for Genomics and Bioinformatics, University of California, Irvine, CA, USA 
 }
%
%
\maketitle

{\footnotesize Originally presented at ``The World is a Jigsaw: Tessellations in the Sciences,'' Lorentz Center, Leiden, The Netherlands, March 2006.}

\section{Introduction}
\label{sec:1}

The shoot apical meristem (SAM) is a dome-shaped
 collection of cells at the apex of growing plants from which all above-ground tissue ultimately derives.  In \textit{Arabidopsis thaliana}  (thale cress), a  small flowering weed of the \textit{Brassicaceae} family (related to mustard and cabbage), the SAM typically contains some three to five hundred cells that range from five to ten microns in diameter. These cells are organized into several distinct zones that maintain their topological and functional relationships throughout the life of the plant. As the plant grows, organs (primordia) form on its surface flanks in a phyllotactic pattern that develop into new shoots, leaves, and flowers.   Cross-sections through the meristem reveal a pattern of polygonal tessellation that is suggestive of Voronoi diagrams derived from the centroids of cellular nuclei.  In this chapter we explore some of the properties of these patterns within the meristem and explore the applicability of simple, standard mathematical models of their geometry. 
 
 \section {Meristem Structure}
\begin{figure}
\centering
\includegraphics[width=10cm]{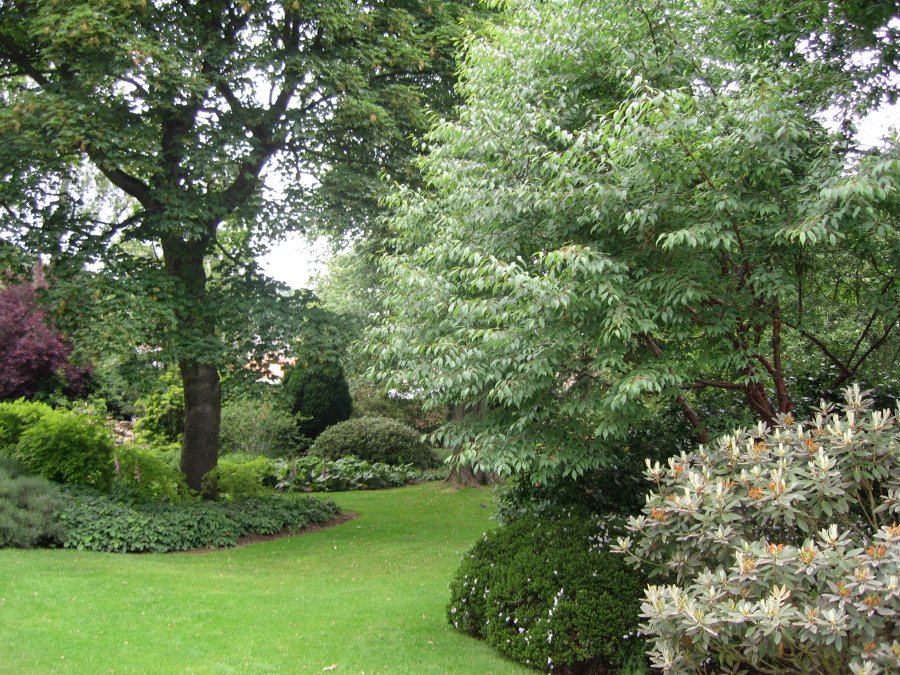}
\caption{All land plants, ranging from the shortest grasses to the tallest trees, have meristems. }
\label{fig:garden}       
\end{figure}

\begin{figure}
\centering
\includegraphics[width=11cm]{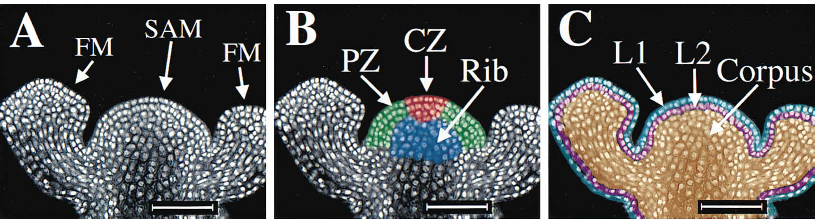}
\caption{A. Laser scanning confocal microscope optical section of SAM and adjacent floral meristems (FM) of \textit{Arabidopsis}. B. Approximation locations of Peripheral Zone (PZ), Central Zone (CZ) and Rib Meristem (Rib). C. Epidermal (L1), Sub-epidermal (L2), and Corpus (L3) Layers. The scale bar is 50 microns. From \cite{Cell97}. }
\label{fig:Cell}       
\end{figure}

All land plants, ranging from the tiny \textit{Arabidopsis} to the tallest sequoia trees, have meristems at the tips of their growing shoots (figure \ref{fig:garden}). This region contains pluripotent stem cells that continue to divide and differentiate into mature tissue throughout the life of the plant.  Meristems maintain their morphological structure and continue to form throughout the life of the organism. In flowering plants this dome-shaped structure can be generally divided into three distinct zones (figure \ref{fig:Cell}):
\begin{itemize}
\item The Central Zone (CZ) on the surface of the dome, surrounding its apex; cell division in the central zone contributes to maintenance of the meristem. It is distinguished by the presence the small (12 amino acids in length) signalling peptide CLAVATA3\cite{CLV3}. 
\item The Peripheral Zone (PZ), also on the surface, but surrounding the apex;  cell division in the PZ contributes to the formation of new organs following the activation of specific biochemical pathways. The PZ is distinguished by the activation of the UFO (for Unusual Floral Organs) gene, whose presence is required for the proper development of flowers\cite{UFO}.
\item The Rib Meristem, deeper within the dome, beneath the CZ; cell division here and in the PZ  results in the elongation of the shoot and leads to the entire structure moving up and leaving older, differentiated cells behind in the shoot. The rib meristem is distinguished by the presence of CLAVATA1, a transmembrane receptor protein whose ligand may be CLAVATA3\cite{CLV1}; and WUSCHEL, a homeodomain transcription factor\cite{WUS}. 
\end{itemize}
 
The majority of cell division occurs in the peripheral zone and rib meristem, with slower division occurring in the central zone.
The exact borders of these zones are not rigorously defined, but instead are distinguishable by the lineages derived thereof (CZ: meristem cells; PZ: leaves, flowers, and shoots; Rib: elongation of the stem). The relative size of each region varies between organisms, but the two surface layers (CZ, PZ) are typically two to three cells thick, and the central zone is typically comprised of five to ten cells surrounding the apex of the meristem \cite{COPB,Cell97}.\\

Meristems are also organized into two or three clonal layers, depending on the species, labeled L1, L2, and L3\cite{Satina40}. Clonal groups are populations of cells that are derived from the same ancestral line established after early embryonic cell divisions.  The surface or outer layer (L1) of cells typically contains epidermal cell precursors. These cells ultimately differentiate into the outer surface layers of the plant. Cells in this layer show a unique pattern of cell division. As cells divide, new cell walls that are added are almost exclusively perpendicular to the outer surface of the SAM. This pattern is known as anticlinal division.

The first sub-epidermal layer (L2) also typically shows anticlinal divisions but the pattern is not always so exclusive.   Sub-epidermal tissue throughout plant organs tend to be derived from the L2 layer. Beneath these two layers (some plants only have a single layer) lies the corpus, or body of the meristem, also designated as L3. Cell division is quite different within the corpus, showing no regular pattern, with the cell walls visually appearing to occur in random directions. The central cells of organs, such as leaves, are derived from the L3 layer.  \\

We have found that the pattern of organization can be maintained in a growth model that stipulates four mutually interacting genes, representing each of the four regions PZ, CZ, Rib, and Stem, where ``Stem'' represents differentiated tissue beneath the meristem \cite{BGRS04}. These genes can be thought of as hypothetical markers that identify each of the four regions. Each gene is assumed to be active and self-activating in its home region, and mutually inhibitory of all of the other markers.  Interactions are represented by a system of neural network equations \cite{JTB91, Cellerator, Vohradsky}. In this representation we represent the important molecular players in the signalling and control network (typically these are proteins or other gene products) by a vector of concentrations $\mathbf{v}$. We refer to these chemical species  $v_i$ collectively as ``genes'' even though in fact each vector component usually represent some immediate downstream product of a gene, such as a protein. In this case we define four ``genes'' that correspond to the observable protein concentrations in each region, with $v_1,v_2,v_3$, and $v_4$ representing the PZ, CZ, Rib, and Stem protein concentrations, respectively. 
The  time evolution of the concentration of $v_i$ in cell $p$, written as $v_i^{(p)}$, is given by\cite{JTB91}
\begin{equation}
\label{eq:nnet}
\frac{dv_i^{(p)}}{dt}=\frac{1}{\tau_i}g\left(u_i^{(p)}\right)
\end{equation}
where
\begin{equation}
\label{eq:nnetnet}
u_i^{(p)}=\sum_{p\in \textnormal{Nbrs}(p)}\sum_j T_{ij}v_j^{(p)} + h_i
\end{equation}
is the net input to any node in the network 
and $g(x)$ is any saturating sigmoidal function such as 
\begin{equation}
\label{eq:sigmoid}
g(x)=\frac{1}{2}\left[1+\frac{x}{\sqrt{1+x^2}} \right]
\end{equation}
The set $\textnormal{Nbrs}(p)$ gives the indices of all the cells that are neighbors of cell $p$, that is, share a common wall with cell $p$. 
The model is then determined by the parameters $\tau_i$, $T_{ij}$, and $h_i$, which must be tuned to reproduce the desired behavior.  A positive value for $T_{ij}$ indicates that when $v_j$ is present, the quantity of $v_i$ will tend to increase; this corresponds to activation. Similarly, a negative value of $T_{ij}$ indicates that the presence of $v_j$ leads to a decrease in the rate of production of $v_i$, or an inhibitory influence.  In the simplest model, species $v_i$ and $v_j$ are allowed to interact with one another if they are either in the same cell, or in an adjoining cell.  In a more complex model one could include a specified rate of diffusion between cells as well as more sophisticated (e.g., biochemical) descriptions of each interaction. Such interactions would lead to additional terms added to the right hand side of equation \ref{eq:nnet}.\\
  \begin{figure}
\centering
\includegraphics[width=9cm]{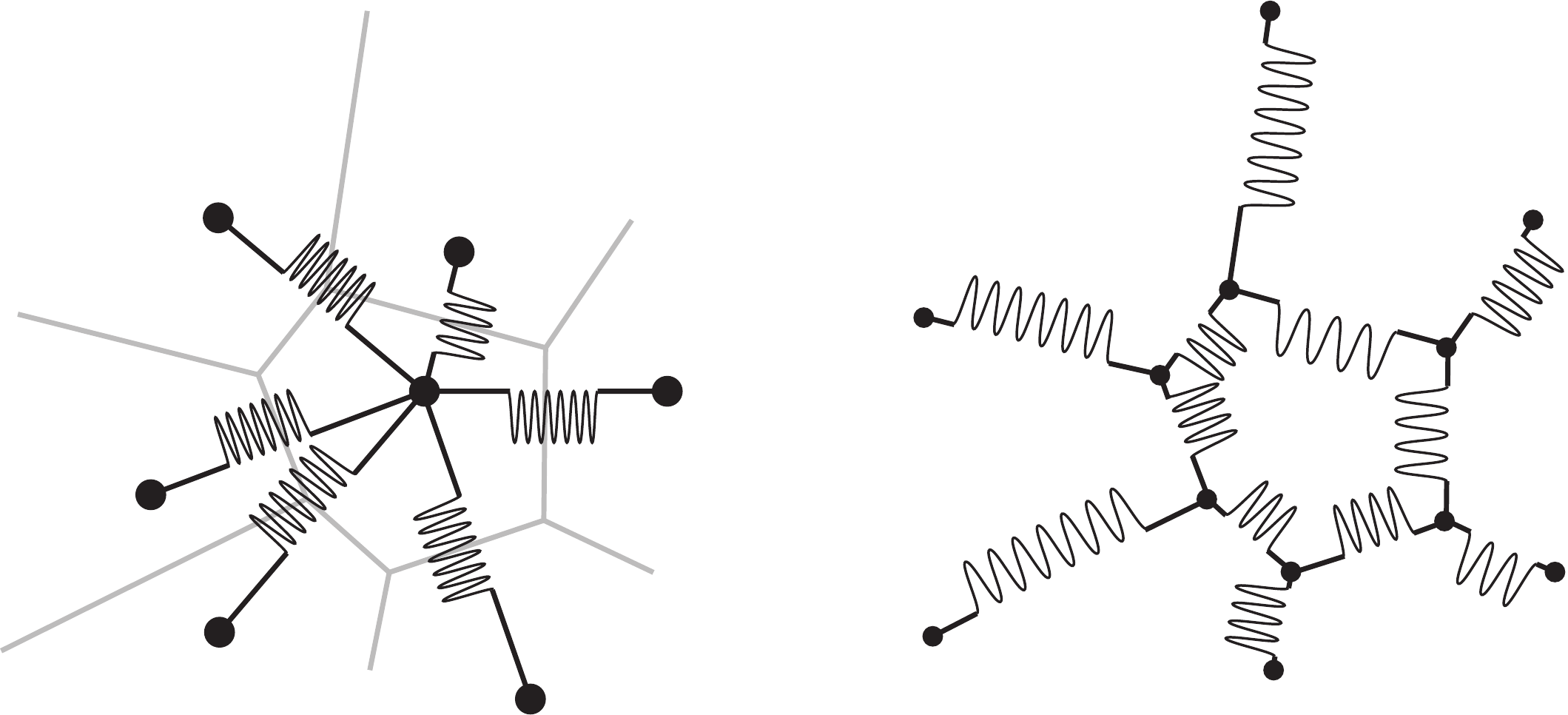}
\caption{Left. In the ``spring model'' tissue is represented in a simple graph structure in which the vertices represent cells (each with an associated mass value) and the links represent cell-cell interactions (each with an associated spring). The total potential on a given cell (equation \ref{eq:springpotential}) is minimized by gradient descent to simulate cell dynamics.  Right. In an alternative ``wall-spring'' model, nodes in the graph represent cell vertices, and links correspond to cell walls with attached springs. In both figures cells are represented by their Voronoi visualization and the springs by their Delaunay links. However, there is no guarantee that either of these  dynamics will maintain either the Voronoi walls or  the Delaunay connectivity. The model on the left is appropriate when only the cell centers are known but there is no a-priori knowledge giving the locations of the cell walls. The model on the right is more general because if the original locations of the cell walls are known it can be generated without any need to calculate Voronoi diagrams.   }
\label{fig:spring-model}       
\end{figure}

\begin{figure}
\centering
\includegraphics[width=10cm]{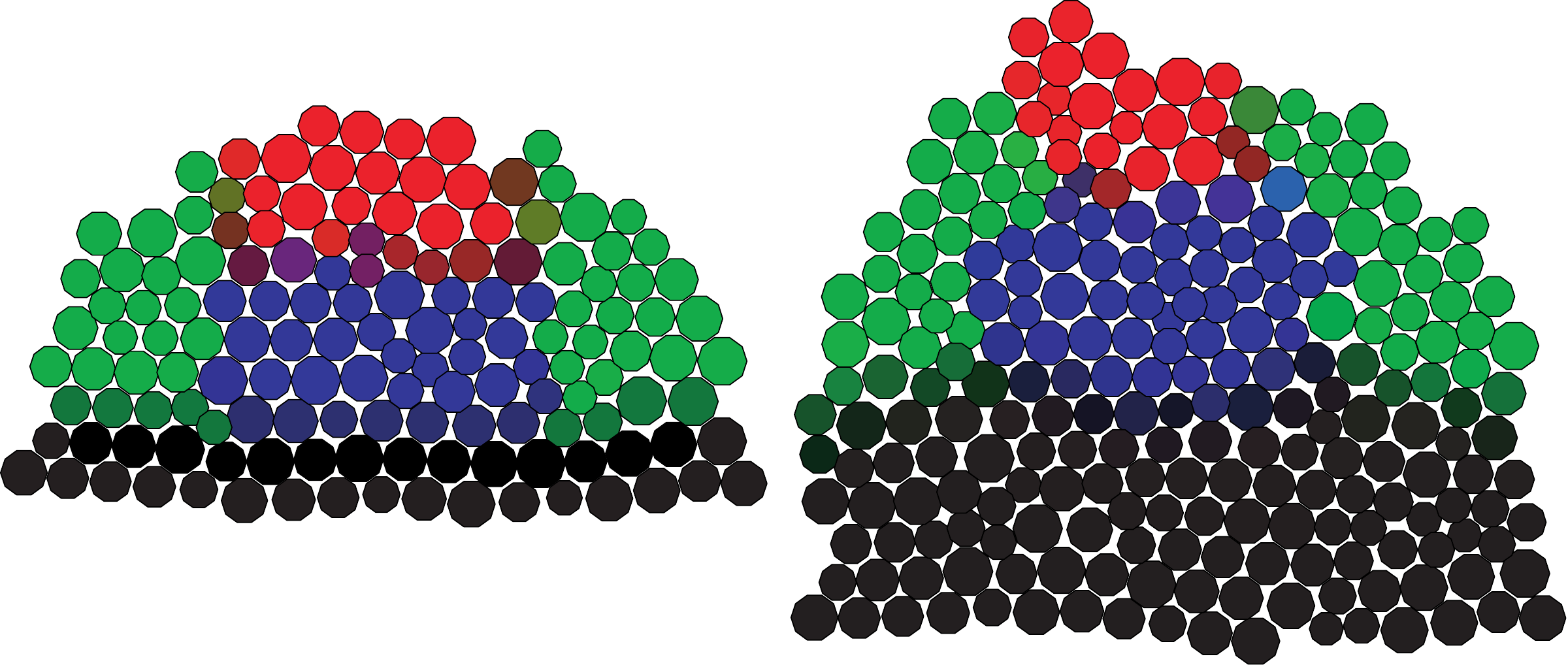}
\caption{Result of a simulation based on four self-activating, mutually-inhibitory species at two arbitrary time points in the simulation. The meristem is segmented into four functions regions, representing the CZ (red), PZ (green), rib meristem (blue), and STEM (black).  }
\label{fig:BGRS}       
\end{figure}

A simple graph structure is used to represent the tissue, as illustrated in figure \ref{fig:spring-model} \cite{BGRS04, Cellerator}. The nodes of the graph represent cells and the links represent cell-cell interactions. 
Each cell is represented by a point mass $m_p$ at position $\mathbf{x_p}$. Nuclear coordinates are then determined by a spring-mass model, in which the point masses are affixed to the cell centers, and each link is represented by a ``spring.'' The dynamics of cell $p$ is represented by simultaneously minimizing the ``spring potentials''
\begin{equation}
\label{eq:springpotential}
V_p=\frac{1}{2}\sum_{q\in\textnormal{Nbrs}(p)}k_{pq}c_{pq}\left[ 
\left(\left| \vec{x}_p - \vec{x}_q\right| - \ell_{pq}\right)^2 - \kappa
\right]
\end{equation}
for all $V_p$.  Here $c_{pq}=1$ when $\ell_{pq}\le d$ and $c_{pq}=0$ when $\ell_{pq}>d$. The $k_{pq}$ are the ``spring constants'',  $\ell_{pj}$ is the equilibrium distance between the point masses as at $\vec{x}_p$ and $\vec{x}_q$, $\kappa$ is a constant that describes the depth of the potential well, and $d$ is a constant distance beyond which cells are assumed to no longer have any mechanical interaction.  Cell growth is then modeled by increasing the equilibrium distances according to $(d/dt)(\ell_{pq})=\mu \ell_{pq}$ for some fixed constant $\mu$, mimicking the observation that cells tend to grow a rates proportional to their masses. A cell continues to grow in this manner until it passes a set threshold volume, in which case it divide asymmetrically into two daughter cells. The volume (mass) of this cell is computed by a spherical approximation with diameter equal to the average distance to neighboring cells. The direction of cell division is random but preferentially in the vertical direction so that the meristem elongates. 
At each cell division the two daughter cells are initially located a small vertical distance apart, with the mass approximately split in half. In practice the minimization can be implemented by performing a gradient descent on equation \ref{eq:springpotential}, e.g., $d\vec{x}_p/dt=-\nabla V_{p}$.  This is akin to an exponential relaxation towards local minimum of a spring in a highly viscous medium. This local minimization technique converges each vertex to its nearest stable equilibrium, is more likely to be physically reachable than a global optimization of the summed potential, and is a solution of stiff-spring dynamics in a viscous media.  The results of a typical simulation is visualized in figure \ref{fig:BGRS}, which shows the tissue at two different time points.\\

In an alternative spring-mass network model (first proposed by Lindenmayer \cite{ABOP}) the  actual cell geometry is represented by a network of interconnected springs. The nodes of this graph network correspond to the actual positions of the cell vertices, and the links are  cell walls. Each link has a spring constant associated with it and dynamics are similarly described by increasing the spring resting length, with the cell walls ``rearranging'' themselves to minimize the forces on each vertex. In addition to spring forces, an outward pointing  pressure force proportional to the length of each cell wall, but applied at the vertices, is required for system stability\cite{Rudge}.  Cell mass can now be computed as the actual area of each polygonal region, which no longer necessarily corresponds to a Voronoi cell. In
Rudge's implementation  \cite{Rudge} cell walls are added by inserting new walls through the cell center aligned in the direction of a randomly chosen polarity vector associated with each cell, which then flips by 90 degrees after each cell division. As with the model depicted in figure \ref{fig:spring-model} this means that cell growth must occur at a much slower rate than the rate at which these forces propagate through the cell well.  This model has been shown to produce convincing geometric representations of plant organ shape.  But also like the spring and mass model described above, it does not provide a realistic model of either the actual physical forces acting on the cells, the actual patterns of cell division, or the genetic program that controls meristem growth. 

\section{Tessellations in simple plant growth models with cell divisions}

Several proposal for generating plant-like two-dimensional tissues
exist (e.~g.~\cite{nakielski,smith, Rudge}). These models are based on global
spatial tissue growth where new walls are added following a set of specific
rules while keeping the cell sizes finite. When  the growth functions are properly tuned, these models can produce very plant-like cell tessellations even though the  mechanical properties of the walls are not included in the model.\\

\begin{figure}
\centering
 \includegraphics[width=11cm]{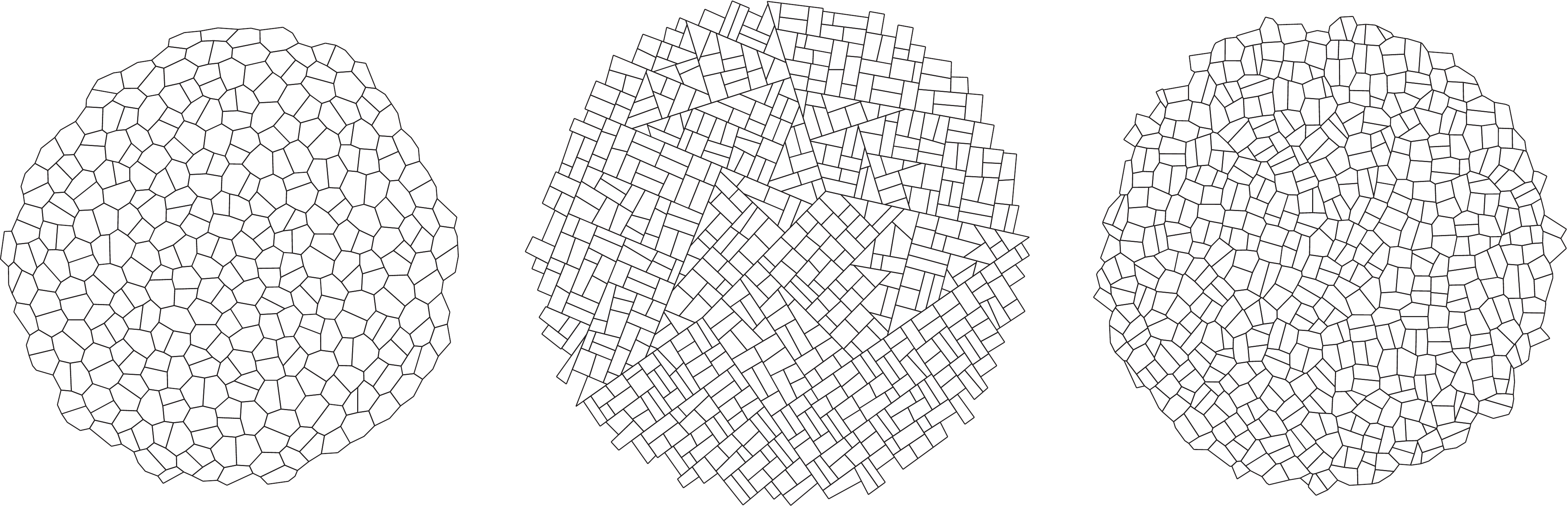}
  \caption{Simulations of global growth implemented by moving
   vertices outwards in the radial direction. Left: initial conditions. Center: result at a later time point of a simulation that does not include any mechanical wall properties. Right: result at a later time point of a simulation that includes mechanical wall properties. }
  \label{mechSim}
\end{figure}

To study the effects of wall mechanics in such models we introduced a simple radial growth into a two-dimensional wall-spring model according to
\begin{equation}
\frac{d\vec{x}_p}{dt} = \nu  \vec{x}_p,
\end{equation}
where $\vec{x}_p$ is the position of vertex $p$ and $\nu$ is a
constant determining the rate of growth. The walls are represented as
springs in a viscous medium as described in equation \ref{eq:springpotential}. In addition, plastic
growth of the walls is introduced by increasing the spring resting
lengths according to
\begin{equation}
\frac{d\ell_{pq}}{dt} = \alpha \Phi(|\mathbf{x_p}-\mathbf{x_q}|-\ell_{pq}) 
\end{equation}
where $\alpha$ is a constant and $\ell_{pq}$ is the natural spring length. $\Phi$ is a linear ramp function, $\Phi(x)=x$ if $x\geq 0$
and $\Phi(x)=0$ if $x<0$. As cells reach a threshold size they divide by introducing a new wall perpendicular to (and centered on) the longest wall. The opposite wall
is identified and the new wall is not allowed to connect to a vertex
(in which case it is moved slightly on the opposite wall). Figure \ref{mechSim} shows two simulations starting at the same
initial condition, but where the mechanical wall properties were only
applied in one (the figure on the far right). The simulations include multiple cell
generations and cells are removed when their radial positions are
outside a maximal value.

These simple simulations indicate that mechanical properties
of walls are important for cell shapes in a growing tissue. The simulation using wall mechanics keeps the shapes of the
cells fairly symmetric and resembles what is seen in the epidermal
layer of the meristem (see figure \ref{fig:zstack}). \\

Plant growth is probably regulated by internal turgor pressure in
individual cells, and the global growth rule used here is a very crude
approximation of this. Also microfibrils in the cell walls introduce a
mechanical asymmetry in most plant cells, although this asymmetry is
not very prominent at the very apex of the SAM. To complicate things
even more, there is feedback between growth and mechanics, and gene
products can affect both growth rates and mechanics. The conclusion
is that much new knowledge is needed before the irregular
tessellations at the plant shoot can be fully understood, and modeling
might be crucial for investigating the global behavior of different
local hypotheses.

 \section{Voronoi Model of Cell Shape}
In the lab we (MH, AR, EMM)  observe several cell type-specific fluorescent markers for growth and differentiation in live \textit{Arabidopsis} plants with a dedicated confocal laser scanning microscope. These markers are affixed to various gene products or promoter regions using green fluorescent protein (GFP) variants that fluoresce when they are illuminated within the microscope by a laser of appropriate wavelength.  This
allows us to observe various meristem and floral primordial features, such as  membranes and nuclei, and to track specific cell lineages over time.  By fitting computational models to these spatiotemporal expression patterns, we can infer 
 how primordial cells are progressively specified and organs develop. From this we develop forward simulations and visualizations of the growing SAM.\\
 \begin{figure}
\centering
\includegraphics[width=11cm]{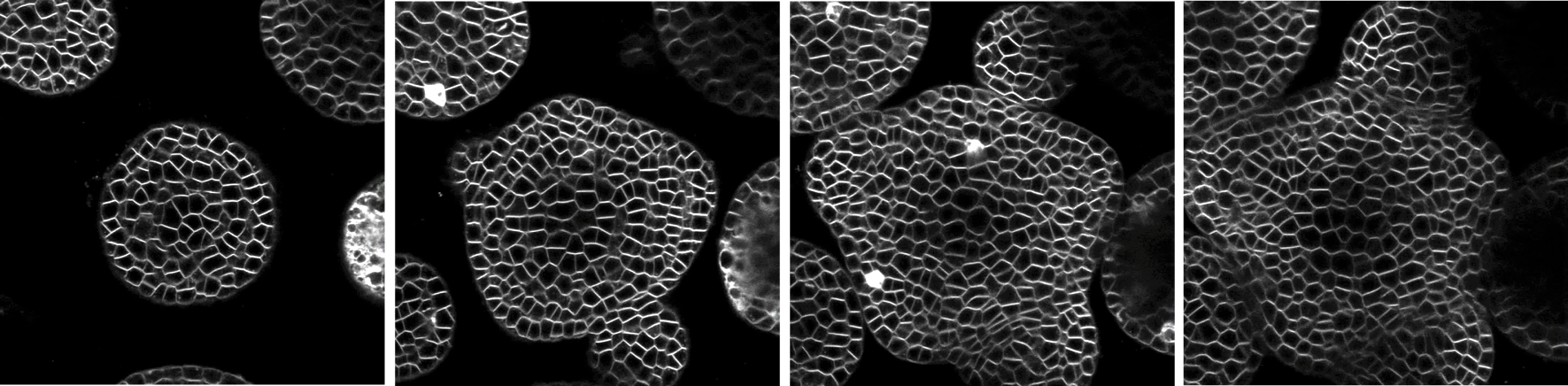}
\caption{Representative images of an \textit{Arabidopsis} SAM from a z-stack showing a membrane bound marker. Every 10th image from a stack of 51 images is shown.  }
\label{fig:zstack}       
\end{figure}
 
  \begin{figure}
\centering
\includegraphics[width=10cm]{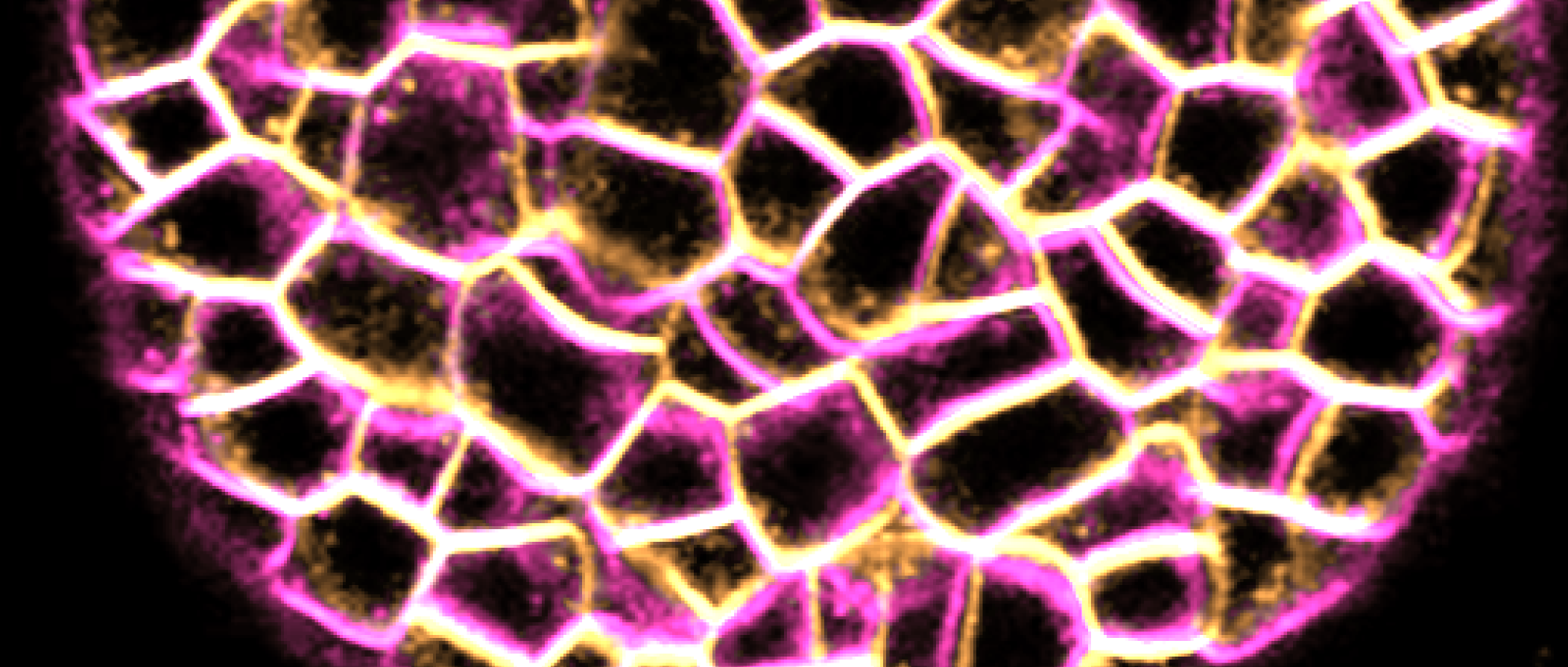}
\caption{Portions of two consecutive sections of an \textit{Arabidopsis} SAM in the same z-stack are overlaid in false colors. In some cases it is easy to determine by inspection which cells correspond to one another.  }
\label{fig:overlay}       
\end{figure}

The three-dimensional (3D) reconstruction starts from stacks of horizontal cross sections that show cell walls and/or subcellular regions marked with different colors of dyes (figure \ref{fig:zstack}). 
These sections are combined to produce four-dimensional visualizations (three spatial dimensions plus time). In any 3D image stack there is a correspondence problem: which cells in one image correspond to which cells in the adjacent cross-section? (See figure \ref{fig:overlay}.)  Cell membranes that are transverse to the image are clearly visible, but it is possible to miss nearly horizontal walls that lie between sections. Such horizontal or nearly horizontal walls must therefore be inferred by the image analysis algorithm. With a time-course of 3D stacks, the formation of floral meristems, cell growth,  and division all complicate the problem.  A further complication is introduced by the fact that the plants must be physically removed from the microscope between time points, and successive images have to be registered to compensate for positional differences that may occur as a result.\\
   
It is not always possible to observe the cell walls in an image. In most cases only one or two different proteins are tagged with GFP and it is extremely difficult to increase the number of markers. Furthermore, adding more tags increases the likelihood of damage to the plant.  Often the markers will not be located in the cell walls but in the nucleus or cytoplasm, as illustrated in figure \ref{fig:nuclei}. While it is not possible to segment the image into individual cells based on wall data, which is not observed in these plants,  it is still desirable to segment the images into different regions that correspond to individual cells for further analysis. In order to fit predictive mechanistic descriptions of cell shape we would prefer that this segmentation match as closely as possible the true cell shape. Often these nuclear markers will show up as diffuse bright spots with a point of maximum intensity that can be used to specify the centroid of the cell's nucleus. These centroids can be used as Voronoi centers for image segmentation.
Even when many of the cell wells can be observed, due to noisy data or other influences it is not always possible to identify all cell wells.  In particular, the exterior walls of epidermal cells and cell walls that are perpendicular to the scan direction (e.g., parallel horizontal sections)  will not be observed.  This led us to explore the hypothesis that meristematic cell walls may be described by the Voronoi diagram of the nuclear centers, this led us to  propose using a Voronoi description for the following tasks:
\begin{figure}
\centering
\includegraphics[width=10cm]{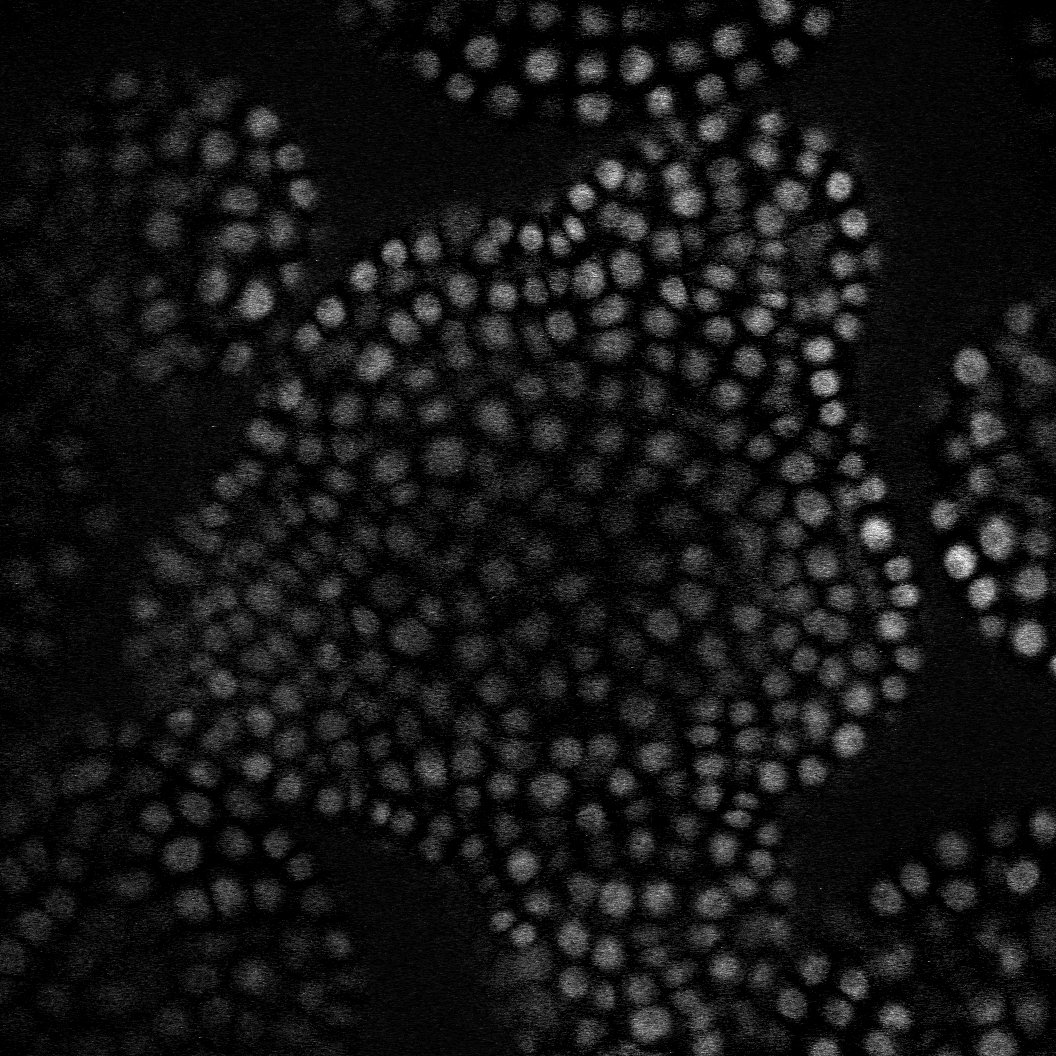}
\caption{Meristem cross section with nuclear marker.}
\label{fig:nuclei}
\end{figure}
\begin{enumerate}
\item Approximation to the location of cell walls when only cell centers are known;
\item First guess approximation to the location of cell walls when only limited or noisy wall data is available;
\item Approximation (or first guess) to the location of outer tissue edge;
\item The location of cell walls that are (nearly) transverse to the imaging angle, and hence may not appear on any of the image slices.
\end{enumerate}
The true Voronoi diagram is certainly not sufficient to accomplish these tasks because Voronoi cells on the edge of the image may either project to infinity, or may exhibit spiky projections that far exceed the true boundary of the tissue (see, for example, figure \ref{fig:voronoiAlgorithm}B).  Even those cells that do not have these properties still tend to have odd shapes which do not match well with the actual tissue boundaries. Instead, we define a truncated Voronoi diagram, in which the outer parts of these cells are chopped off. Construction of the truncated Voronoi diagram is illustrated in two dimensions in figure  \ref{fig:voronoiAlgorithm}, and proceeds as follows: 

\begin{figure}
\centering
\includegraphics[width=11 cm]{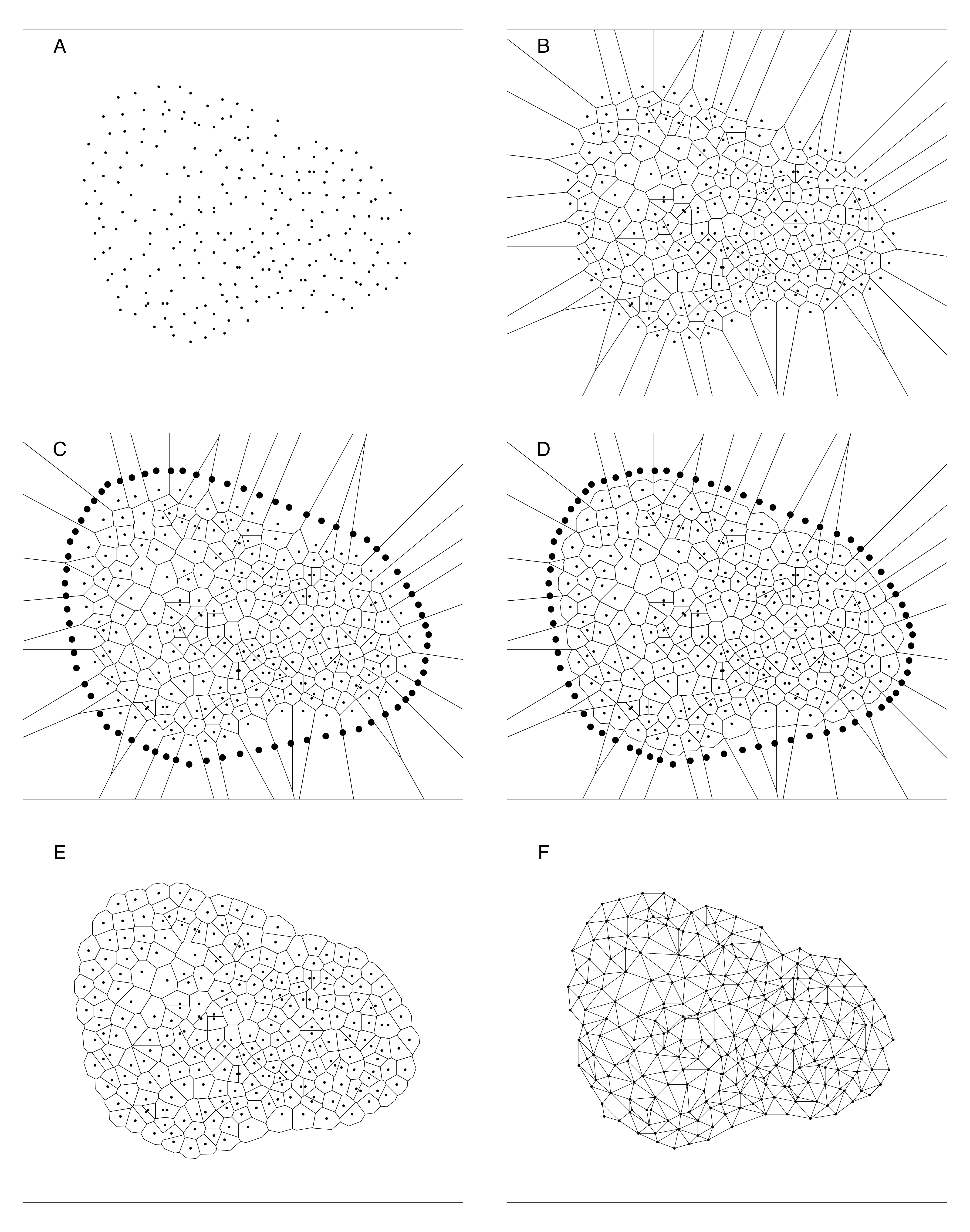}
\caption{Construction of the truncated Voronoi diagram. (A) Initial two-dimensional data set, giving nuclear locations. (B) Full Voronoi diagram based on initial data; Voronoi cells on the edge do not realistically approximate the tissue boundary. (C) Addition of artificial points around the convex hull. (D) Construction of Voronoi diagram of original plus added cells. For clarity, the Voronoi walls of only the original cells are shown, not the added cells. (E) All Voronoi cells and centers corresponding to the added cells are dropped, to give the final visualization. (F) The Delaunay Triangulation of the resulting visualization gives an accurate connectivity diagram.}
\label{fig:voronoiAlgorithm}       
\end{figure}

\begin{enumerate}
\item Input the list of cell centers in Cartesian coordinates.
\item Calculate the convex hull of the cell centers.
\item Calculate the Delaunay triangulation of the cell centers.
\item Calculate $r_{av}$, the average length of a link the Delaunay triangulation -- this represents the average cell diameter.
\item Define an extended convex hull by projecting the convex hull perpendicularly outward from the tissue by a distance $(1+\epsilon)r_{av}$ where $\epsilon$ is a small parameter that is tuned to optimize the appearance of the visualization.
\item Define a layer of artificial cell centers along the extended convex hull. In general these need to be placed somewhat more closely together than $r_{av}$ to prevent cells from ``leaking'' outward in step \ref{item:VoronoiCalculation} - typically about $r_{av}/2$ apart is close enough  (figure \ref{fig:voronoiAlgorithm}C).
\item Define an extended cell set as the union of the true cell centers and the artificial cells defined in the prior step.
\item \label{item:VoronoiCalculation} Calculate the Voronoi diagram of the extended cell set  (figure \ref{fig:voronoiAlgorithm}D).
\item Remove the Voronoi cells that correspond to the artificial cells; what remains is the Voronoi tissue approximation (figure \ref{fig:voronoiAlgorithm}E).
\item Calculate the Delaunay triangulation of the extended cell set (true plus artificial centers). Any cells whose Delaunay connections include one end at an artificial center and one end at a true center are epidermal (L1) cells. 
\end{enumerate}
\begin{figure}[h]
\centering
\includegraphics[width=11cm]{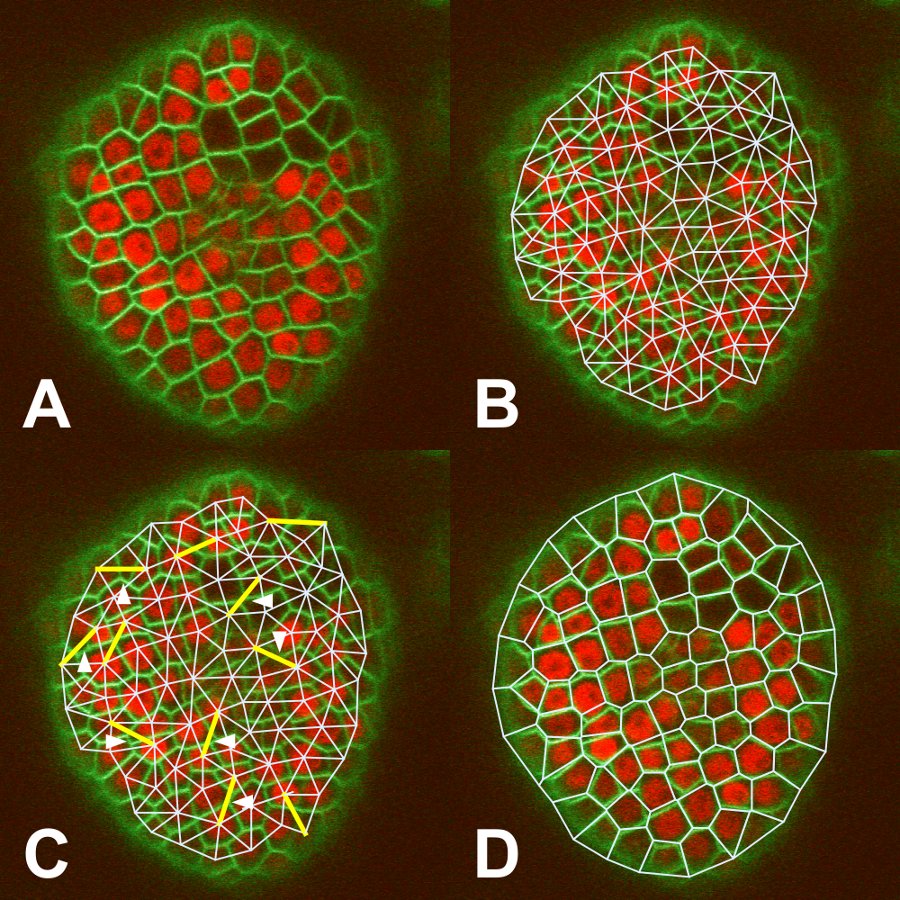}
\caption{(A) GFP cross-sectional image of \textit{Arabidopsis} SAM showing nuclear (red) and cell wall (green) markers. A gradient ascent to maximum intensity was used to identify nuclear centers; these centers were then used to compute the two-dimensional tessellations shown in B-D. This section is the 14th layer down from the top of a 64-image stack at 2.25$\mu$ spacing. (B) Same image with the pruned Delaunay connections superimposed. (C) Delaunay links that do not correspond to cell connections are highlighted (11 of 233 Delaunay links). The arrows indicate Delaunay cell connections that are not Delaunay links (7 of 229 actual cell-cell interfaces). (D) The corresponding truncated  Voronoi of same image with the facets removed using $\epsilon=0.3$. }
\label{fig:voronoiData}       
\end{figure}
An application of this algorithm is illustrated in figure \ref{fig:voronoiData}, which shows the truncated Voronoi diagram and Delaunay links overlaid on top of a true image. 
Given that no compelling biological data was used to determine these wall locations, the Voronoi boundaries match the physical cell wall locations surprisingly well.   Furthermore the Delaunay triangulation identifies which cells directly interact with one another quite nicely, even when the cell walls themselves do not quite match up. For example, in the section illustrated in figure \ref{fig:voronoiData}, there are 229 cell-cell interfaces between 88 cells. Of these 229 actual interfaces, the pruned Delaunay algorithm correctly identified 222 links (96.9\%). The pruned Delaunay triangulation of this section produced 233 links, including the 222 correct links and 11 ``false'' links.  Some error is to be expected in this type of calculation because although the image corresponds to a single horizontal cross-section the nuclei themselves do not. In a few cases the ``false'' links can either be attributed to portions of cells for which nuclear data is not available, or to the rare presence of 4-way cellular junctions. Because of the nature of confocal microscopy, any such image will also include some light from cells slightly above or below the plane of the image. Furthermore, the cell nuclei themselves extend through several image planes and their actual centroids will rarely lie on the exact image plane. The same effect would occur by examining the meristem surface, although in this case all of the nuclei would lie below the image plane, and there would be the added complication of calculating a Voronoi on a curved surface.\\

\begin{figure}
\centering
\includegraphics[width=11 cm]{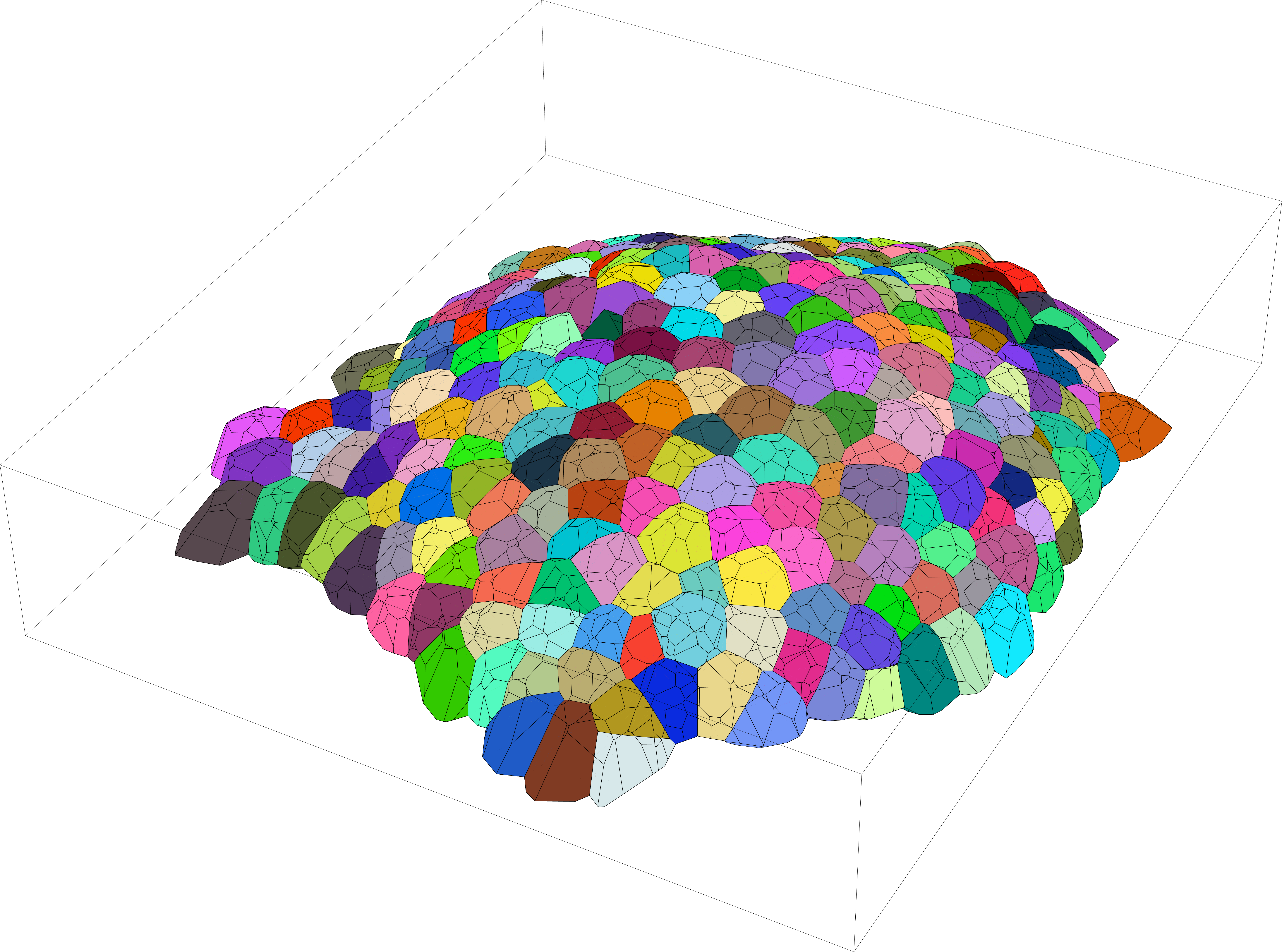}
\caption{Example of the truncated Voronoi diagram in three dimensions.}
\label{fig:voronoiAlgorithm3}       
\end{figure}

It is possible to improve the appearance of the faceted look (figure \ref{fig:voronoiAlgorithm}E) by using a smoothing algorithm or by modifying the outer edge of the visualization. For example, one could increase the value used by $\epsilon$ and then throw away all vertices that  ``stick out,'' e.g., any vertex on the outer edge that does not include an interior link as well as an edge link. The problem with this is that it removes all of the facets, and in fact some of the facets should be there. A closer inspection of the image will reveal areas where small portions of cells on other z-levels project into the edge and therefore should be included in the visualization, but are in fact not included because their nuclei are missing from the image.  These small segments of cells would naturally smooth out the space between two adjacent cells. Furthermore there are sometimes other cells in the interior of the image that are also missed by the algorithm, for the same reason. Such cells have nuclei that lie too far outside the focal plane of the image to actually be seen in the given section. The weakness here would appear to be due to the loss of information that occurs as  a result of truncating the data from three to two dimensions.  This problem might presumably be fixed by utilizing a three-dimensional Voronoi diagram. This algorithm can be easily extended to three dimensions; a typical result is illustrated in figure \ref{fig:voronoiAlgorithm3}.   \\

 \begin{figure}
\centering
\includegraphics[width=8cm]{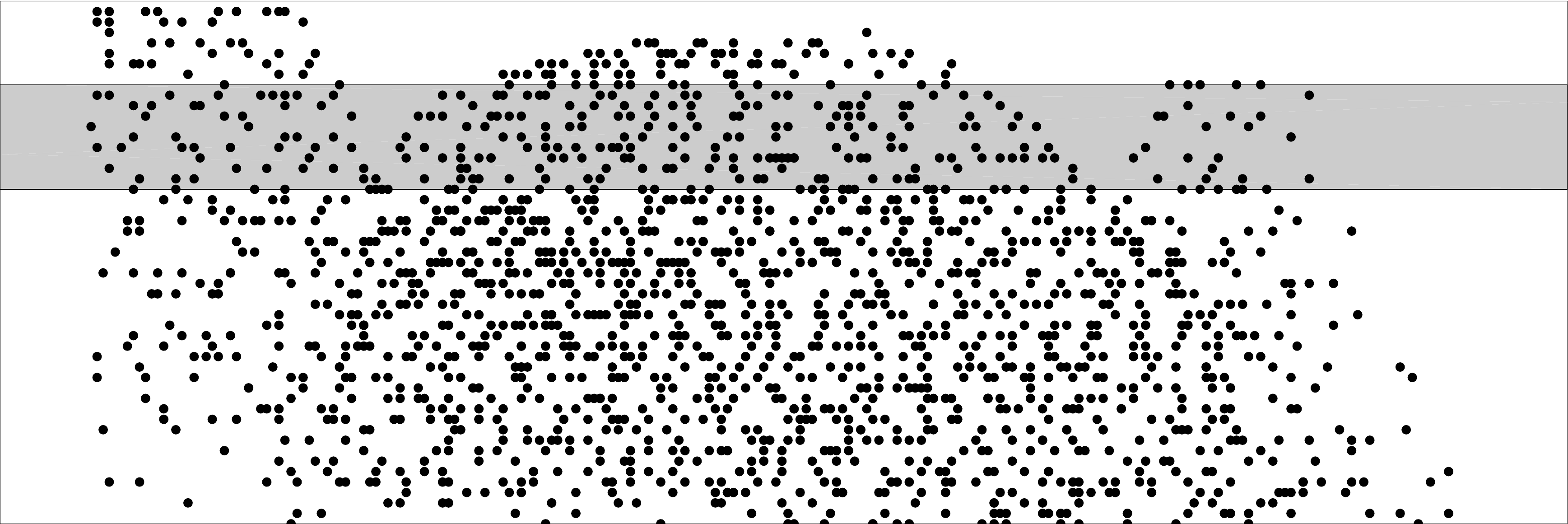}
\caption{Projection onto a vertical plane of data set containing 1872 cell centers. The grayed area shows the 10-micron slab at z=67 microns that is compressed to generate a two-dimensional section. The image clearly shows the large dome-shaped shoot apical meristem flanked by two floral meristems. The dimensions of the figure are 150 microns (horizontal) by 50 microns (vertical).  }
\label{fig:Slabbing}       
\end{figure}

\begin{figure}
\centering
\includegraphics[width=8cm]{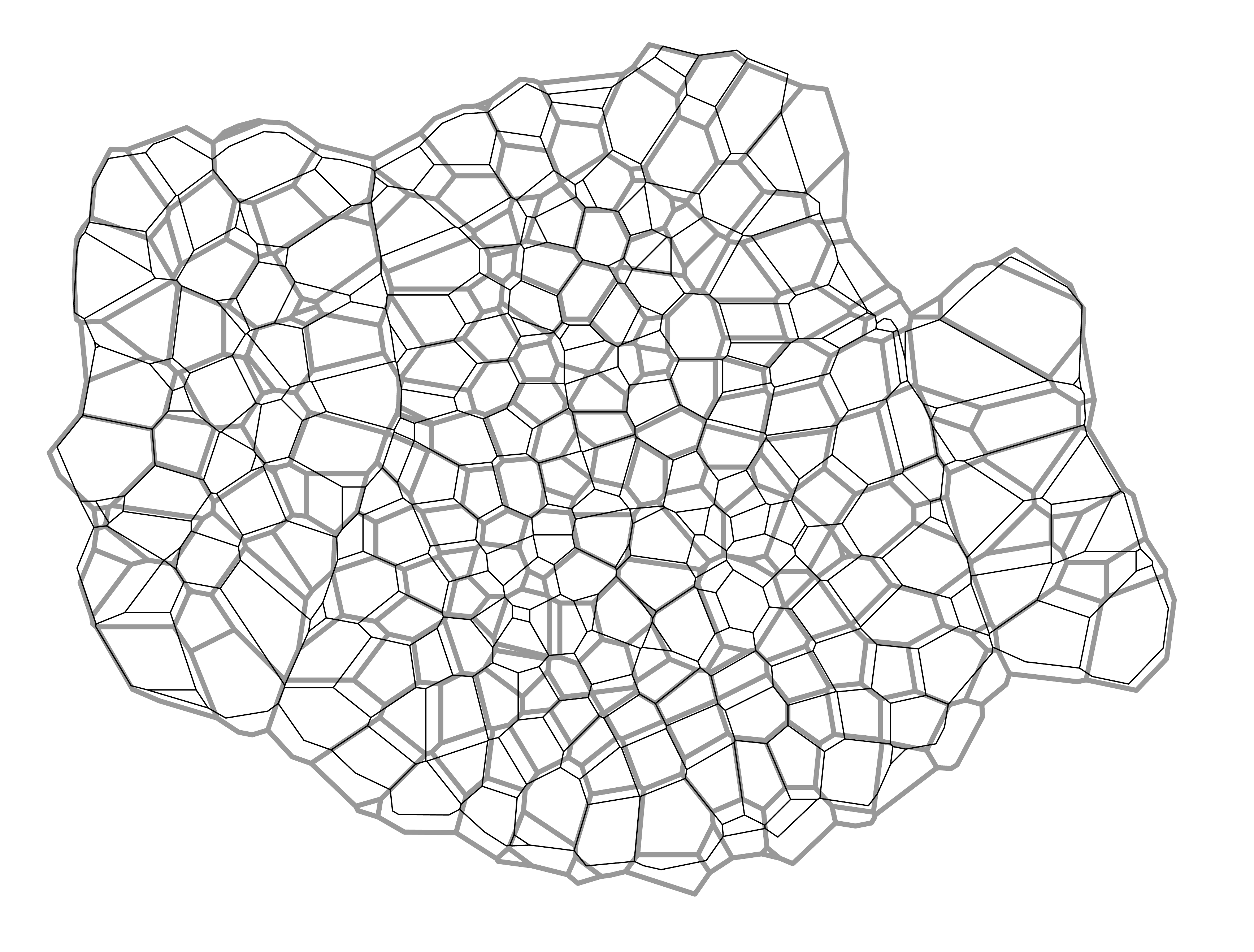}
\caption{Horizontal planar sections of Voronoi visualization at z=72 microns (thin lines) and z=70 microns (thicker lines).}
\label{fig:72vs73}       
\end{figure}

\begin{figure}
\centering
\includegraphics[width=8cm]{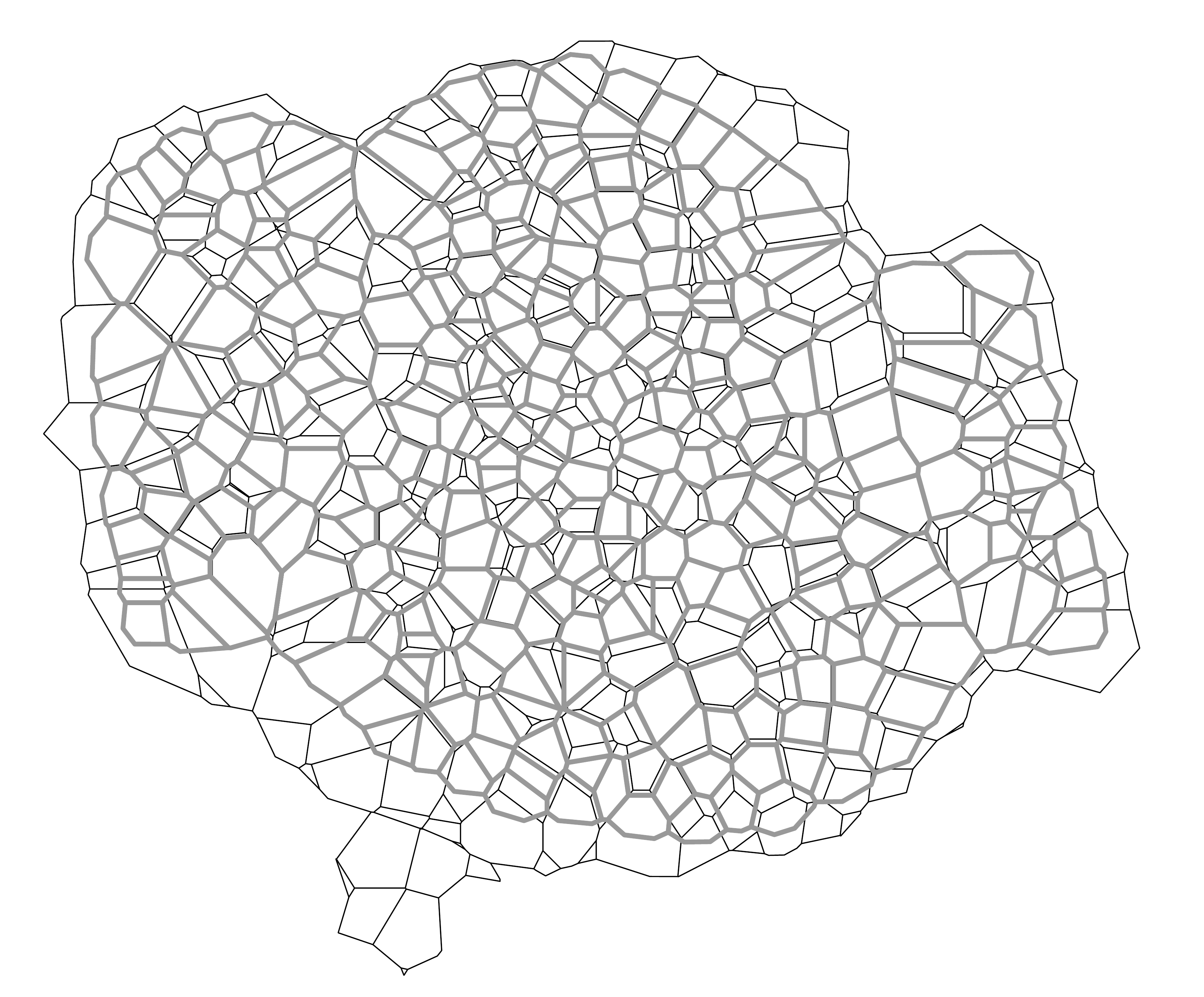}
\caption{Comparison of planar section of three-dimensional Voronoi visualization at z=67 microns (thin lines) with two-dimensional Voronoi visualization of a 10 micron thick slab centered at 67 microns (thicker lines).  Some cells that are visible in the three-dimensional cross section, such as the appendage at the bottom, are completely missing from the two-dimensional slab visualization, because no nuclei appear in the images.}
\label{fig:TwoDVsThreeD}       
\end{figure}

If the Voronoi hypothesis were correct, it would seem that a better estimate of cell wall locations would be provided by calculating actual cross-sections of three-dimensional Voronoi diagrams rather than calculating a two-dimensional Voronoi diagram from a single focal plane. The difficulty is illustrated in figure \ref{fig:Slabbing}. The fluorescence image shows not just nuclei that exist in the actual focus plane but parts of nuclei that exist both above and below the focus plane.  Data observed at a single focal plane thus really includes some information about cells within a ``slab'' of finite thickness rather than an infinitely thin layer. Furthermore, the observed cell nuclei in the images are not actually points but extended ``blobs'' whose z-coordinates are less accurately known then their xy-coordinates. The actual coordinates used are the centroids of these blobs determined by a gradient descent algorithm on the image intensity. Thus any image will show not a true cross-section but a flattened projection of data within a finitely thick slab of the meristem. Finally, there will be some cells whose nuclei can not be seen at all, because the nuclei are either completely above or completely below the plane of visualization, but which have portions of the cell that intersect the plane of focus. Consequently it is not obvious which horizontal cross-section should actually be used in looking at a particular image (figures \ref{fig:72vs73} and \ref{fig:TwoDVsThreeD}). 
 \begin{figure}
\centering
\includegraphics[width=8cm]{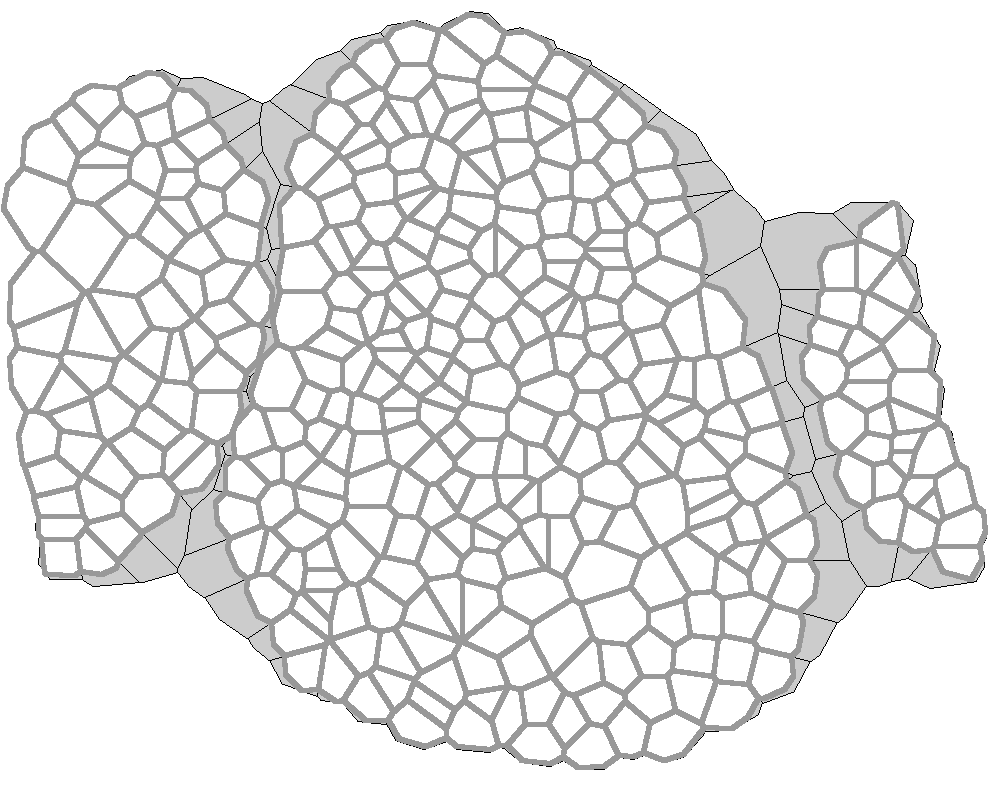}
\caption{Segmentation of shoot apical meristem from floral meristems. }
\label{fig:FloralSegments}       
\end{figure}
A second problem with the Voronoi method is that it only works for convex or nearly convex data; it does not automatically segment different regions that are actually physically disconnected. For example, in the slab displayed in figure \ref{fig:Slabbing} we see that there are actually three separate regions that will appear in the data: the central shoot apical meristem and the two flanking floral meristems. A simple application of the truncated Voronoi visualization will determine that there are (unnaturally large) cells in the crevices between the three segments (see the gray cells in figure \ref{fig:FloralSegments}). In the data illustrated, the three regions were segmented manually and then recombined to produce the visualization with the white cells in figure \ref{fig:FloralSegments}. It is difficult to determine an appropriate heuristic that will correctly separate these segments, so it must be done manually.\\

 \begin{figure}
\centering
\includegraphics[width=10cm]{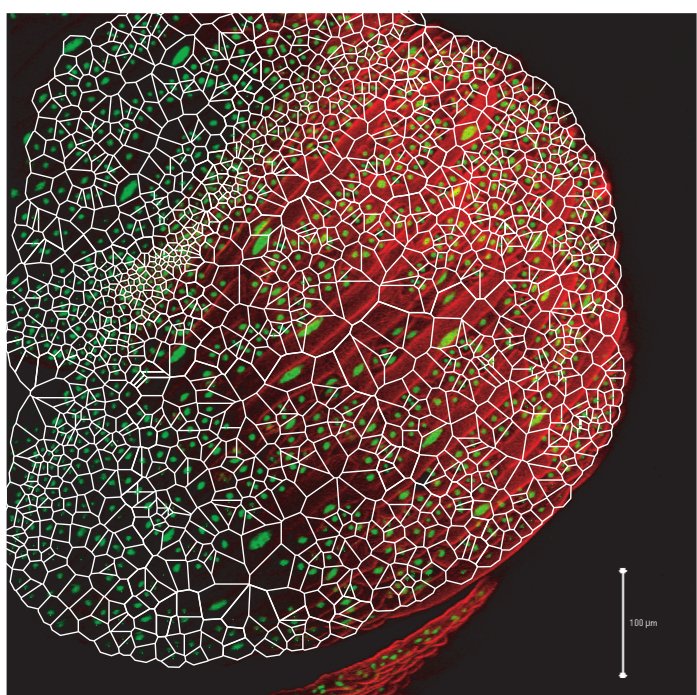}
\caption{Laser scanning confocal image of \textit{Arabidopsis} sepals  shows cell membranes (reddish) and nuclei (green). The corresponding Voronoi diagram is shown in white. }
\label{fig:sepals}       
\end{figure}

It is important to emphasize that while Voronoi-like tessellations may reasonably describe some portions of plant morphology, other parts of the plant show radically different cellular patterns. For example, cells in \textit{Arabidopsis} sepals tend to be elongated along a specified axis and vary greatly in size (figure \ref{fig:sepals}). The sepals are the outermost, green, leaf-like organs that protect and cover the flower while it is developing.   One peculiar property of this organ is the presence of numerous ``giant cells'' in the epidermis -- cells that have grown to enormous size compared to the adjacent cells. The cell nuclei also appear to be growing in size; genetic material appears to be duplicating with the normal cell cycle but new cell walls are not forming, so the cells and their nuclei continue to grow. There appear to be several distinct populations of cells of different sizes, but with the common property that they are all elongated in a common direction. The purpose of this elongation is not known, and it may be structural or somehow related to the mechanical process in which the sepals open to form the flower.  Giant cells are not present in all plant flowers, and their function is also not known. The Voronoi diagram derived from the nuclear centers of the cells in the sepals is also illustrated in \ref{fig:sepals}. As can be seen from this illustration, both the directional elongation and the presence of giant cells contribute to the poor correspondence of the Voronoi diagram to the actual cell walls. While it might be possible to \textit{fit} a Voronoi model to the observations using some sort of generalized Voronoi algorithm \cite{MH05}, power diagrams \cite{Poupon}, or a medial axis transforms \cite{Chromosome}, it seems unlikely that any of these methods will be able to \textit{predict} the cell wall locations based only on the locations of nuclear centers without some additional a-priori knowledge regarding the type of cells and their properties. 

\section{Discussion}
The question of how a plant forms new cell walls has been studied for over a century and a half.
While turgor pressure probably plays a significant role in determining the planes of cell division and ultimately the final cellular tessellation, the actual mechanics of the process are still not known. 
 Three observations were made  in the mid-19th century that could be used to predict the planes of cell division based on purely geometric rules:
\begin{itemize}
\item Hofmeister's Rule (1863): New cell walls are usually formed in a plane perpendicular to the principal axis of cell growth.
\item Sach's rule (1878): New cell walls form in a plane perpendicular to existing cell walls. 
\item Errera's Rule (1888): The plane of division corresponds to the shortest path that will halve the volume of the mother cell.
\end{itemize}
More recently it has been observed that cell division planes tend to be staggered, like bricks in a wall \cite{Sinnott1941}, and in particular, that walls avoid four-way junctions, i.e., it is unlikely that more than three cell walls will intersect at any given point \cite{Lloyd}.    \\

Others \cite{Lynch} have observed that if single cells are compressed they tend to divide in a plane perpendicular to the long axis of the cell.   These observations would seem to indicate a mechanical basis for cell wall placement. This would imply that there are ways the cell can detect forces and respond to them. Other internal and external cues may also affect the process. In particular, it is known that the process of cell division is quite different in plants than in animals, and that the process in plants is probably influenced by mechanical forces. In animal cells the plane of cell division is determined by the position of the mitotic spindle during mitosis. In plants, however,  the plane of cell division is determined during an earlier phase in the cell cycle. During prophase, which initiates mitosis, actin filaments and microtubules align themselves in plant cells, forming a structure known as the pre-prophase band (PPB). The ultimate plane of cell division will be aligned along the PPB.  A structure known as a phragmoplast forms along the PPB, eventually forming the cell plate that divides the daughter nuclei during a later phase of mitosis, and finally forms the cell wall.  Actin filaments and micro-tubules are known to be involved in supporting the mechanical structure of the cell and thus the polarization of the PPB may occur in response to specific mechanical stressors.\\

We have found that Voronoi diagrams are  quite useful for one particular purpose, namely, as an initial guess to the location of cell walls in two-dimensional cross-sections of shoot apical meristems, particularly when only partial or fuzzy data is available. In this application, both cell nuclear and wall locations will be available in the data, and a standard image processing algorithm can be used to estimate the locations of the cell walls. However, if some portions of the cell walls are not identifiable, the Voronoi diagram can be used as a first approximation for the location of the cell walls, and then the cell walls moved around to find an optimal location that best matches the remaining data. Typical image processing algorithms will only give pixel sets or possibly finite element approximations to those detailed pixel sets but not higher level polyhedral face and edge information. No  satisfactory robust algorithm has yet been published to obtain this information even in two dimensions and an enormous amount of manual intervention is still required.  To obtain the polyhedral approximation in three dimensions one typically needs to perform some sort of plane fitting to proposed edges or start with some first guess and then move the vertices around in response to an optimization algorithm. This method will not work, however, if the cells are not polyhedral or if the Voronoi diagram is a bad approximation  (as was the case in figure \ref{fig:sepals}, which is not part of the shoot apical meristem). One possible solution to this has been proposed by \cite{Burl} which obtains a triangulated mesh. \\

One can make reasonable arguments both for and against the Voronoi hypothesis in modeling. In its favor would be the optimal distribution of resources; this is an application of the classical post-office problem. In this argument, cell walls would form in a manner so that the nearest nucleus to any point in the cytoplasm would be a cell's own nucleus. Thus nuclear signals could be optimally distributed.  It is not clear how this organization would be maintained as the plant grows and cell volumes increase, however, and it seems unlikely that cell walls would significantly rearrange themselves after being established. There is also considerable evidence that mechanical forces are a driving impetus for the formation of new cell walls. In this case the near-match of the Voronoi and true cell walls would then be nothing more than a coincidence. \\

The biochemical processes that control cell division and how they are influenced by the various internal stresses and strains that cells experience is a fundamental problem of developmental biology whose solution has yet to be unravelled. There is some indication that theses forces directly influence the changing shape of the meristem, but as yet there is no direct evidence linking this influence to any particular geometrical tessellation.  A wide variety of cellular tessellations have been identified in nature, and purely mechanical models have been developed that predict a continuum of possible stable configurations\cite{Farhadifar}, very few of which correspond to pure Voronoi tessellations, although these models do not incorporate information about the nuclear locations or formation of new cell walls. The problem is complicated by the fact that computation of such tessellations under dynamically changing conditions with a realistic number of cells ($\approx 1000$) is a computationally challenging problem that will probably require the use of high performance computing and massive parallelization. It remains to be seen if a complete mechanical model that takes into account all known stresses and strains upon plant cell walls as well as regulatory feedback mechanisms can provide a set of physical conditions under which the Voronoi or a generalized Voronoi tessellation is expected.
 Such a model would come a long way towards our understanding of the processes that influence plant growth.

\end{document}